\documentstyle[manuscript,aps,prl]{revtex}
\input{epsf}

\begin{document}
\draft

\title{Quantum optical coherence tomography with dispersion cancellation}

\author{Ayman F. Abouraddy, Magued B. Nasr, Bahaa E. A. Saleh, Alexander V. Sergienko, and Malvin C. Teich
\footnote{teich@bu.edu}}

\address{Quantum Imaging Laboratory
\footnote{http://www.bu.edu/qil}, Departments of Electrical $\&$
Computer Engineering and Physics, Boston University, Boston, MA
02215}

\maketitle

\begin{abstract}
We propose a new technique, called quantum optical coherence
tomography (QOCT), for carrying out tomographic measurements with
dispersion-cancelled resolution. The technique can also be used to
extract the frequency-dependent refractive index of the medium.
QOCT makes use of a two-photon interferometer in which a swept
delay permits a coincidence interferogram to be traced. The
technique bears a resemblance to classical optical coherence
tomography (OCT). However, it makes use of a nonclassical
entangled twin-photon light source that permits measurements to be
made at depths greater than those accessible via OCT, which
suffers from the deleterious effects of sample dispersion. Aside
from the dispersion cancellation, QOCT offers higher sensitivity
than OCT as well as an enhancement of resolution by a factor of 2
for the same source bandwidth. QOCT and OCT are compared using an
idealized sample.
\end{abstract}


\section{Introduction}

\noindent Optical Coherence Tomography (OCT) has become a
versatile and useful tool in biophotonics \cite{OCT}. It is a form
of range-finding that makes use of the second-order coherence
properties of a classical optical source \cite{ClassicalOptics} to
effectively section a reflective sample with a resolution governed
by the coherence length of the source. OCT therefore makes use of
sources of short coherence length (and consequently broad
spectrum), such as superluminous LEDs and ultrashort-pulsed
lasers.

A number of non-classical (quantum) sources of light have been
developed over the past several decades \cite{Teich90} and it is
natural to inquire whether making use of any of these might be
advantageous. The answer turns out to be in the affirmative.
Spontaneous parametric down-conversion (SPDC) \cite{SPDC} is a
nonlinear process that generates entangled beams of light; these
have been utilized to demonstrate a number of two-photon
interference effects \cite{TwoPhotonInterference1} that cannot be
observed using traditional light sources
\cite{TwoPhotonInterference2}. We demonstrate that such
entangled-photon fourth-order interference effects may be used to
carry out range measurements similar to those currently obtained
using classical OCT, but with the added advantage of even-order
dispersion cancellation \cite{DispersionCancellation}. This is
possible by virtue of the non-classical nature of the light
produced by SPDC. We refer to this new technique as quantum
optical coherence tomography (QOCT).

\section{Classical Optical Coherence Tomography (OCT)}

\noindent The sample investigated in the course of our
calculations, classical and quantum alike, is represented by a
transfer function $H$. This quantity describes the overall
reflection from all structures that comprise the sample. For an
incident undepleted monochromatic plane wave of angular frequency
$\omega$,
\begin{equation}\label{ContinuousSample}
H(\omega)=\int_{0}^{\infty}dz\:
r(z,\omega)e^{i2\varphi(z,\omega)}.
\end{equation}
Here $r(z,\omega)$ is the complex reflection coefficient from
depth $z$ and $\varphi(z,\omega)$ is the phase accumulated by the
wave while travelling through the sample to the depth $z$.

The basic scheme of OCT \cite{OCT} is illustrated in Fig. 1. We
assume that the classical source produces cw incoherent light with
a short coherence time of the order of the inverse of its spectral
width (the results also apply to the case of a pulsed source,
however). We characterize the source S with a power spectral
density $S(\omega_{0}+\Omega)$ where $\omega_{0}$ is its central
angular frequency. The light is divided by a beam splitter into
the two arms of a Michelson interferometer. A variable delay
$\tau$, imparted by a scanning mirror, is placed in the
``reference arm'' while the sample is placed in the ``sample
arm''. The reflected beams are recombined by the beam splitter and
an interferogram $I(\tau)$ is measured:
\begin{equation}\label{ClassicalInterferogram}
I(\tau)\propto\Gamma_{0}+2\textrm{Re}\{\Gamma(\tau)e^{-i\omega_{0}\tau}\}.
\end{equation}
The self interference term $\Gamma_{0}$ and the cross-interference
term $\Gamma(\tau)$ are given by
\begin{equation}\label{ClassicalReferenceTerm}
\Gamma_{0}=\int
d\Omega\,[1+|H(\omega_{0}+\Omega)|^{2}]\,S(\Omega),
\end{equation}
and
\begin{equation}\label{ClassicalInterferenceTerm}
\Gamma(\tau)=\int d\Omega\,
H(\omega_{0}+\Omega)\,S(\Omega)\,e^{-i\Omega\tau}=h_{c}(\tau)\ast
s(\tau),
\end{equation}
respectively, where $h_{c}(\tau)$ is the inverse Fourier transform
of $H(\omega_{0}+\Omega)$ with respect to $\Omega$, and $s(\tau)$
is the correlation function of the source [the inverse Fourier
transform of $S(\Omega)$]. The symbol $\ast$ represents the
convolution operation.

The physical underpinnings of this scheme may be understood by
examining the interference of light propagating in the two paths
created by the beam splitter (Fig. 1). A monochromatic wave of
frequency $\omega_{0}+\Omega$ emitted from S acquires a reflection
coefficient $H(\omega_{0}+\Omega)$ in the sample arm, but only a
phase factor $e^{i(\omega_{0}+\Omega)\tau}$ in the reference arm.
As is clear from Eq. (\ref{ClassicalInterferogram}), the resulting
interferogram includes a self-interference contribution from the
two paths (Eq. \ref{ClassicalReferenceTerm}): a factor of unity
from the reference path and a factor of
$|H(\omega_{0}+\Omega)|^{2}$ from the sample path. The
cross-interference contribution, which resides in Eq.
(\ref{ClassicalInterferogram}), is the product of these two terms,
$H(\omega_{0}+\Omega)$ and $e^{i(\omega_{0}+\Omega)\tau}$ (one is
conjugated, but this is of no significance in OCT). This term may
also be expressed as a convolution of the sample reflection with
the coherence function of the source, the temporal width of which
serves to limit the resolution of OCT.

\section{Quantum Optical Coherence Tomography (QOCT)}

\noindent The scheme we propose for QOCT is illustrated in Fig. 2.
The twin-photon source is characterized by a frequency-entangled
state given by \cite{EntangledState}
\begin{equation}\label{QuantumState}
|\Psi\rangle=\int d\Omega\,
\zeta(\Omega)\,|\omega_{0}+\Omega,\omega_{0}-\Omega\rangle,
\end{equation}
where $\Omega$ is the angular frequency deviation about the
central angular frequency $\omega_{0}$ of the twin-photon wave
packet, $\zeta(\Omega)$ is the spectral probability amplitude, and
the spectral distribution of the wave packet
$S(\Omega)=|\zeta(\Omega)|^{2}$ is normalized such that $\int
d\Omega\,S(\Omega)=1$. For simplicity, we assume $S$ is a
symmetric function and that both photons reside in a common single
spatial and polarization mode.

Interferometry is implemented by making use of a seminal
two-photon interference experiment, that of Hong, Ou, and Mandel
(HOM) \cite{HOM}. The HOM beam-splitter interferometer is modified
by placing a reflective sample in one of the paths in the
interferometer and a temporal delay $\tau$ is inserted in the
other path, as shown in Fig. 2. The two photons, represented by
beams 1 and 2, are then directed to the two input ports of a
symmetric beam splitter. Beams 3 and 4, the outputs of the beam
splitter, are directed to two single-photon-counting detectors,
D$_1$ and D$_2$. The coincidences of photons arriving at the two
detectors are recorded within a time window determined by a
coincidence circuit. The delay $\tau$ is swept and the coincidence
rate $C(\tau)$ is monitored. If a mirror were to replace the
sample, sweeping the delay would trace out a dip in the
coincidence rate whose minimum would occur at equal overall path
lengths, which we define as zero delay. This dip is a result of
quantum interference of the two photons within a pair.

For a sample described by $H(\omega)$, as provided in Eq.
(\ref{ContinuousSample}), the coincidence rate $C(\tau)$ is given
by
\begin{equation}\label{QuantumInterferogram}
C(\tau)\propto\Lambda_{0}-\textrm{Re}\{\Lambda(2\tau)\},
\end{equation}
where the self-interference term $\Lambda_{0}$ and the
cross-interference term $\Lambda(\tau)$ are defined as follows:
\begin{equation}\label{QuantumReferenceTerm}
\Lambda_{0}=\int d\Omega\,|H(\omega_{0}+\Omega)|^{2}\,S(\Omega),
\end{equation}
and
\begin{equation}\label{QuantumInterferenceTerm}
\Lambda(\tau)=\int d\Omega\,
H(\omega_{0}+\Omega)\,H^{*}(\omega_{0}-\Omega)\,S(\Omega)\,e^{-i\Omega\tau}=h_{q}(\tau)\ast
s(\tau).
\end{equation}
Here $h_{q}(\tau)$ is the inverse Fourier transform of
$H_{q}(\Omega)=H(\omega_{0}+\Omega)\,H^{*}(\omega_{0}-\Omega)$
with respect to $\Omega$.

It is important to highlight the distinctions and similarities
between Eqs. (\ref{QuantumInterferogram}),
(\ref{QuantumReferenceTerm}), and (\ref{QuantumInterferenceTerm}),
and Eqs. (\ref{ClassicalInterferogram}),
(\ref{ClassicalReferenceTerm}), and
(\ref{ClassicalInterferenceTerm}). The unity OCT background level
in Eq. (\ref{ClassicalReferenceTerm}) is, fortuitously, absent in
Eq. (\ref{QuantumReferenceTerm}) for QOCT. Moreover, the QOCT
cross-interference term in Eq. (\ref{QuantumInterferenceTerm}) is
related to the reflection from the sample quadratically; the
sample reflection is therefore simultaneously probed at two
frequencies, $\omega_{0}+\Omega$ and $\omega_{0}-\Omega$, in a
multiplicative fashion. Finally, the factor of 2 by which the
delay in the QOCT cross-interference term in Eq.
(\ref{QuantumInterferogram}) is scaled, in comparison to that in
Eq. (\ref{ClassicalInterferogram}) for OCT, leads to an
enhancement of resolution in the former.

A particularly convenient twin-photon source makes use of
spontaneous parametric down-conversion (SPDC) \cite{SPDC}. This
process operates as follows: a monochromatic laser beam of angular
frequency $\omega_{p}$, serving as the pump, is sent to a
second-order nonlinear optical crystal (NLC). Some of the pump
photons disintegrate into pairs of downconverted photons. We
direct our attention to the case in which the photons of the pairs
are emitted in selected different directions (the non-collinear
configuration). Although each of the emitted photons in its own
right has a broad spectrum, by virtue of energy conservation the
sum of the frequencies must always equal $\omega_{p}$. Because of
the narrow spectral width of the sum frequency (which is the same
as the pump frequency), the photons interfere in pairs. But
because of the broadband nature of each of the photons
individually, they serve as a distance-sensitive probe not unlike
the broadband photons in conventional OCT.

\section{Comparison of QOCT and OCT}

\noindent The sample model presented in Eq.
(\ref{ContinuousSample}) may be idealized by representing it as a
discrete summation
\begin{equation}\label{DiscreteSample}
H(\omega)=\sum_{j}\: r_{j}(\omega)\,e^{i2\varphi_{j}(\omega)},
\end{equation}
where the summation index extends over the layers that constitute
the sample. This is a suitable approximation for many biological
samples that are naturally layered, as well as for other samples
that are artificially layered such as semiconductor devices. This
approximation is not essential to the development presented in
this paper, however.

We further assume, without loss of generality, that the dispersion
profile of the media between all surfaces are identical, so that
$\varphi_{j}(\omega)=\beta(\omega)\,z_{j}$, where
$\beta(\omega)=n(\omega)\,\omega/c$ is the wave number at angular
frequency $\omega$, $z_{j}$ is the depth of the $j^{th}$ layer
from the sample surface, $n(\omega)$ is the frequency-dependent
refractive index, and $c$ is the speed of light in vacuum. We
expand $\beta(\omega_{0}+\Omega)$ to second order in $\Omega$:
$\beta(\omega_{0}+\Omega)\approx\beta_{0}+\beta'\Omega+\frac{1}{2}\beta''\Omega^{2}$,
where $\beta'$ is the inverse of the group velocity $v_{0}$ at
$\omega_{0}$, and $\beta''$ represents group velocity dispersion
(GVD) \cite{ClassicalOptics}.

In the case of OCT, using Eqs. (\ref{ClassicalInterferenceTerm})
and (\ref{DiscreteSample}) leads to a cross-interference term
given by
\begin{equation}\label{COCTDiscreteSample}
\Gamma(\tau)=\sum_{j}r_{j}\,s^{(0j)}_{d}\!\!\left(\tau-2\frac{z_{j}}{v_{0}}\right)\,e^{i2\beta_{0}z_{j}},
\end{equation}
where $s_{d}^{(0j)}(\cdot)$ arises from reflection from the
$j^{th}$ layer after suffering sample GVD over a distance
$2z_{j}$, the subscript $d$ indicates dispersion, and the
superscript $(0j)$ indicates that dispersion is included from the
surface of the sample $(0)$ all the way to the $j^{th}$ layer. The
quantity $s_{d}^{(jk)}(\cdot)$ is thus the Fresnel transformation
of $S(\Omega)$ with dispersion coefficient $\beta''$
\cite{ClassicalOptics}:
\begin{equation}\label{FresnelTransformation}
s_{d}^{(jk)}\!(\tau)=\int d\Omega\,
S(\Omega)\,e^{i2\beta''\Omega^{2}(z_{j}-z_{k})}\,e^{-i\Omega\tau}.
\end{equation}
The effectiveness of OCT is therefore limited to samples that do
not exhibit appreciable GVD over the depth of interest.

In the case of QOCT, on the other hand, Eqs.
(\ref{QuantumInterferenceTerm}) and (\ref{DiscreteSample}) result
in a cross-interference term given by the sum of two
contributions:
\begin{equation}\label{QOCTDiscreteSample}
\Lambda(\tau)=\sum_{j}|r_{j}|^{2}\,s\!\left(\tau-4\frac{z_{j}}{v_{0}}\right)+\sum_{j\neq
k}r_{j}\,r^{*}_{k}\,s_{d}^{(jk)}\!\!\left(\tau-2\frac{z_{j}+z_{k}}{v_{0}}\right)\,e^{i2\beta_{0}(z_{j}-z_{k})};
\end{equation}
the first contribution represents reflections from each layer
\textit{without GVD}, while the second contribution represents
cross-terms arising from interference between reflections from
each pair of layers. The quantity $s(\cdot)$ is the correlation
function of the source defined previously, and the quantity
$s_{d}^{(jk)}(\cdot)$ is the Fresnel transformation given in Eq.
(\ref{FresnelTransformation}). In contrast to OCT, only dispersion
between the $j^{th}$ and $k^{th}$ layers survives, as is evident
by the superscript $(jk)$. The terms comprising the first
contribution in Eq. (\ref{QOCTDiscreteSample}) include the
information that is often sought in OCT: characterization of the
depth and reflectance of the layers that constitute the sample.
The terms comprising the second contribution in Eq.
(\ref{QOCTDiscreteSample}) are dispersed due to propagation
through the inter-layer distances $z_{j}-z_{k}$; however, they
carry further information about the sample that is inaccessible
via OCT. Two complementary approaches can be used to extract
information from Eq. (\ref{QOCTDiscreteSample}): 1) averaging the
terms that comprise the second contribution by varying the pump
frequency while registering photon coincidences such that the
exponential function averages to zero, which leads to unambiguous
optical sectioning information resident in the first contribution;
and 2) isolating and identifying the terms of the second
contribution to obtain a more detailed description of the sample
than is possible with OCT. This can, in fact, be achieved by
making use of the Wigner distribution as will be demonstrated at
the end of this section.

We now proceed to provide a numerical comparison between QOCT and
OCT using Eqs. (\ref{COCTDiscreteSample}) and
(\ref{QOCTDiscreteSample}). Consider a sample comprising of two
reflective layers buried at some depth below the surface of a
medium, as illustrated at the very top of Fig. 3. For the purposes
of our calculation, we arbitrarily choose amplitude reflection
coefficients $r_{1}=0.1$ and $r_{2}=0.2$, separation distance
$d_{1}=10\,\mu$m, and depth below the sample surface
$d_{0}=0.1\,$mm. For both OCT and QOCT, calculations are carried
out by assuming that the source has a central wavelength
$\lambda_{0}=2\pi c/\omega_{0}=812\,$nm and a Gaussian spectral
distribution with a bandwidth (full width at $1/e$ of maximum) of
155 nm, which corresponds to a wave packet of temporal width 14
fsec and length 4.2 $\mu$m in free space. In the context of QOCT,
this can be realized by means of a $\beta$-barium borate NLC of
thickness 1 mm pumped by a source of wavelength $\lambda_{p}=406$
nm. Using type-I SPDC, a NLC cut at an angle $29^{\circ}$ with
respect to the optic axis generates light centered about the
degenerate wavelength $\lambda_{0}=2\lambda_{p}=812$ nm. For
purposes of illustration, we neglect reflection from the top
surface of the sample and assume that the sample dispersion
profile is characterized by:
$\beta'=5\times10^{-9}\,\textrm{s}\,\textrm{m}^{-1}$ and
$\beta''=1.8\times10^{-25}\,\textrm{s}^{2}\,\textrm{m}^{-1}$.
These correspond to a highly dispersive material; however, common
materials such as ``higher dispersion crown glass'' \cite{CRC} are
an order-of-magnitude more dispersive.

The results of this calculation are displayed in Fig. 3 for OCT
(thin rapidly varying gray curve) and QOCT (black broken curve
representing the full signal; black solid curve representing the
signal averaged over pump frequency). Because of dispersion it is
clear that no useful information about the sample is available
from OCT. QOCT, on the other hand, yields a pair of
high-resolution dispersion-cancelled coincidence-rate dips at
delays corresponding to reflections from the two surfaces.
Moreover, the QOCT resolution is a factor of 2 superior to that
achievable via OCT in a dispersionless medium. The peak between
the two dips evidenced in the full QOCT signal (black broken
curve), which could alternatively be a dip depending on the phases
of the terms in the second contribution in Eq.
(\ref{QOCTDiscreteSample}), is a result of quantum interference
between the probability amplitudes arising from reflection from
the two different surfaces. This is in contrast to the black
solid-curve dips, which are a result of quantum interference
between the probability amplitudes arising from reflection from
each surface independently. The breadth of the middle peak is
determined only by the dispersion of the medium residing between
the two reflective surfaces and not by the nature of the material
under which they are buried. It is clear, therefore, that the
dispersion of the region between the two surfaces may be
determined by measuring the broadening of the middle peak in
comparison with the two dips.

It is worthy of note that dispersion cancellation occurs for all
even powers of the expansion of $\beta(\omega)$. Thus if the
phases of reflection from the surfaces are random, which provides
a model for transmission through a turbid or turbulent medium,
only the middle peak will wash out, while the dips arising from
reflections from the surfaces of interest are unaffected. In OCT,
such random phase variations serve to deteriorate, and possibly
destroy, information about the sample.

In Fig. 4 we plot results for the same example examined above,
except that one of the layers of interest is situated at the
surface of the sample rather than being buried beneath it. In this
case OCT gives intelligible results although the return from the
second layer is clearly broadened as a result of dispersion. On
the other hand, the results for QOCT are identical to those shown
in Fig. 3 for the same two-layer object buried under a dispersive
medium. QOCT is also seen to exhibit higher sensitivity than OCT
for weakly reflective samples. This is because the
self-interference term in QOCT [Eq. (\ref{QuantumReferenceTerm})]
does not include the factor of unity present in the
self-interference term of OCT [Eq.
(\ref{ClassicalReferenceTerm})].

Finally, we address the use of the Wigner distribution for
extracting information about the sample via the QOCT
cross-interference term $\Lambda$. Examining Eq.
(\ref{QuantumInterferenceTerm}), and assuming that the bandwidth
of $S(\Omega)$ is greater than that of $H_{q}(\Omega)$, we obtain
\begin{equation}\label{WignerDistribution}
\Lambda(\tau,\omega_{0})\approx\int d\Omega\,
H(\omega_{0}+\Omega)\,H^{*}(\omega_{0}-\Omega)\,e^{-i\Omega\tau}.
\end{equation}
This is precisely the Wigner distribution function of the function
$H(\Omega)$, with parameters $\tau$ and $\omega_{0}$
\cite{Wigner}. Knowledge of $\Lambda(\tau,\omega_{0})$ for all
relevant values of $\tau$ and $\omega_{0}$ guarantees that
$H(\Omega)$ may be reconstructed \cite{Cohen}. The quantity $\tau$
is varied by changing the delay in path 1 of the interferometer in
Fig. 2. The quantity $\omega_{0}$ may be changed by varying the
pump frequency $\omega_{p}=2\omega_{0}$. Although this technique
might be expected to face practical difficulties because the
direction of SPDC changes as the frequency is varied, this could
be mitigated by adjoining a wave-guiding mechanism to the
twin-photon source, as is customary when using periodically-poled
NLCs, for example. Furthermore, such an approach would enable the
output light to be directly coupled into an optical fiber and thus
integrated into systems already familiar to the practitioners of
OCT.

\section{Advanced-wave interpretation}

The operation of QOCT may be understood in a heuristic way by
considering an advanced-wave interpretation similar to that
employed by Klyshko in the context of spatial interferometers
\cite{Klyshko}. In such an interpretation one of the detectors may
be thought of as being replaced by a classical light source with
its waves traced backward through the optical system, and the
twin-photon source may be thought of as a reflector. The intensity
measured at the location of the other detector then mimics the
coincidence rate \cite{Klyshko}. Applying this interpretation to
QOCT, assume that D$_{1}$ in Fig. 2 is replaced by a classical
light source that emits a monochromatic wave of frequency
$\omega_{0}+\Omega$. The beam splitter results in this wave being
partitioned into two paths. In one of these (path 1) the wave
travels backward through the delay, changes direction and flips
its frequency about $\omega_{0}$ to $\omega_{0}-\Omega$ at the
twin-photon source, and then propagates forward through path 2.
Finally, it reflects from the sample and reaches D$_{2}$ having
acquired a weighting factor of
$e^{i(\omega_{0}+\Omega)\tau}\,H(\omega_{0}-\Omega)$. The second
wave (path 2 after the beam splitter) reflects from the sample,
changes direction and frequency from $\omega_{0}+\Omega$ to
$\omega_{0}-\Omega$ at the source, and then undergoes a delay
$\tau$ in path 1 en route to D$_{2}$, acquiring a weighting factor
$e^{i(\omega_{0}-\Omega)\tau}\,H(\omega_{0}+\Omega)$. The
self-interference contribution in Eq. (\ref{QuantumReferenceTerm})
is given by the sum of the squared amplitudes of these two terms.
The cross-interference contribution to the interferogram is one of
these terms multiplied by the complex conjugate of the other:
$e^{-i2\Omega\tau}\,H(\omega_{0}+\Omega)\,H^{*}(\omega_{0}-\Omega)$.
This interpretation makes clear the origin of the salutary
time-scaling by a factor of 2 and the absence of interference
fringes at frequency $\omega_{0}$ from the QOCT interferogram.
Both QOCT interfering waves reflect from the sample and they do so
at conjugate frequencies, whereas in OCT one of the waves reflects
from a mirror, which gives rise to the deleterious unity term that
is absent from QOCT. In contrast to OCT, the complex conjugate
present in the cross-interference term is of central importance in
QOCT.

\section{Conclusion}

\noindent We have presented a new technique, called quantum
optical coherence tomography (QOCT), which utilizes the
wave-packet nature of photons generated in pairs via spontaneous
parametric down-conversion (SPDC). Each photon of the pair
inherently occupies a broad spectrum even though the pump is
monochromatic: the bandwidth is determined by the length of the
nonlinear crystal. QOCT yields performance superior to that of a
classical optical coherence tomography (OCT) on three counts: 1)
the resolution is enhanced by a factor of 2 for the same source
bandwidth; 2) it has greater sensitivity for weakly reflecting
samples; and 3) sample group-velocity dispersion does not result
in a deterioration of resolution with increasing depth into the
sample. Moreover, the frequency dependent refractive index of the
medium, which is inaccessible to OCT, may be extracted.

{\em Acknowledgments.---} We thank Zachary Walton for helpful
discussions. This work was supported by the National Science
Foundation and by the Center for Subsurface Sensing and Imaging
Systems (CenSSIS), an NSF engineering research center.

\begin{figure}[h]
 \centering
 \epsfxsize=6.8in \epsfysize=5.2in
 \epsffile{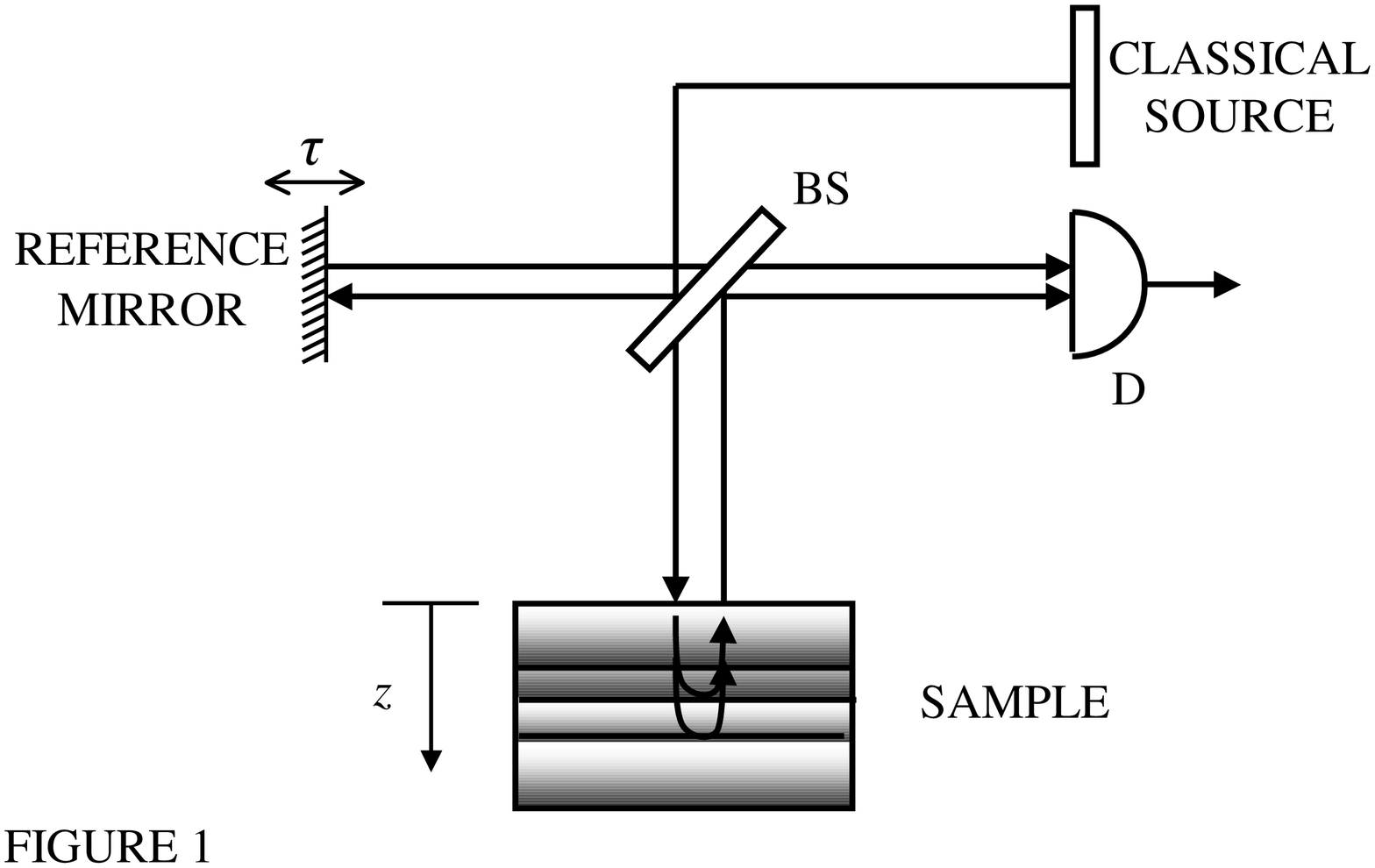}
 \vglue 0.2cm
 \label{Fig.1}
 \caption{Setup for optical coherence tomography
(OCT). BS stands for beam splitter; D is a detector; and $\tau$ is
a temporal delay introduced by moving the reference mirror.}
 \end{figure}

\begin{figure}[h]
 \centering
 \epsfxsize=6.8in \epsfysize=5.2in
 \epsffile{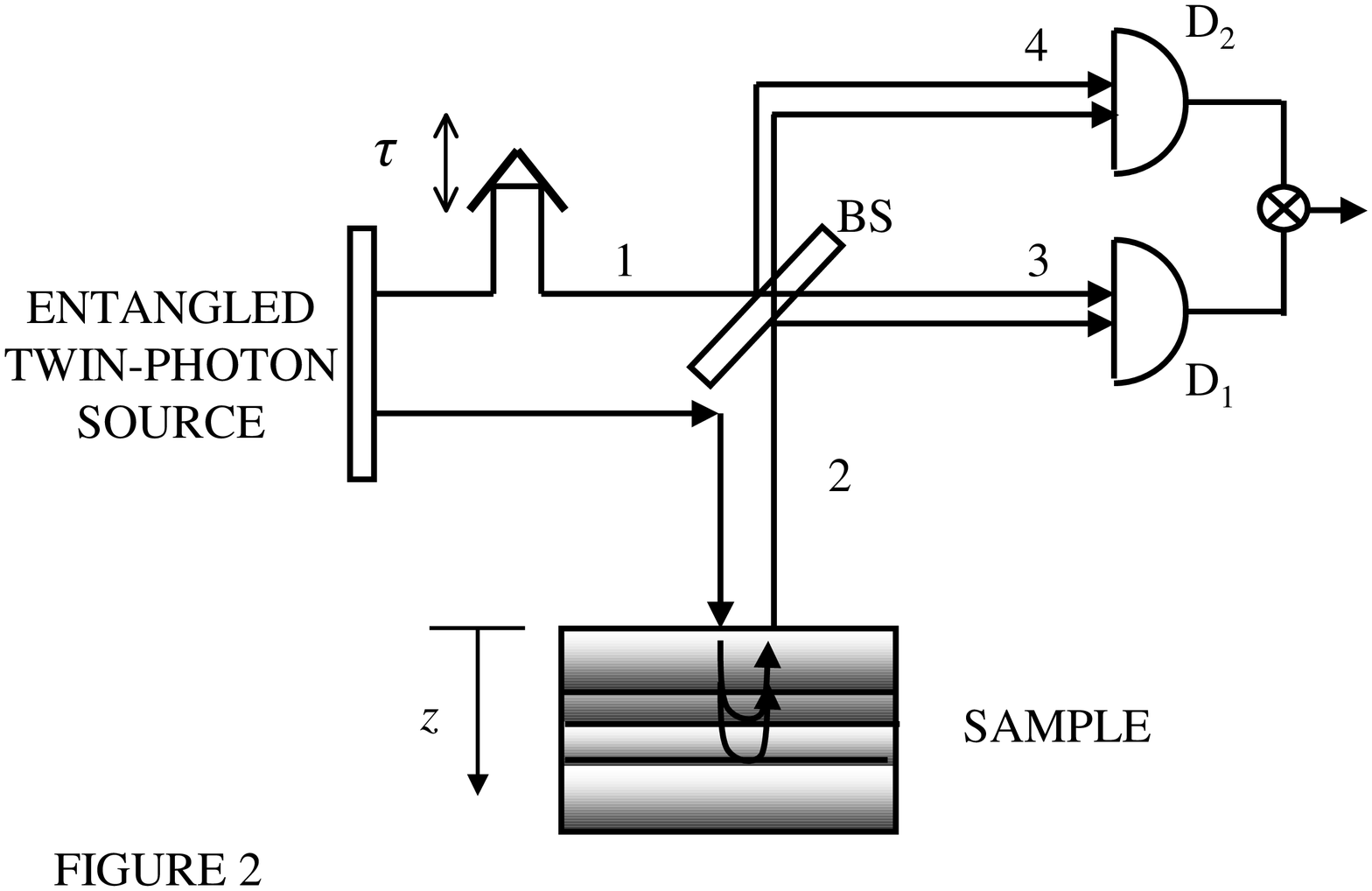}
 \vglue 0.2cm
 \label{Fig.1}\caption{Setup for quantum optical coherence
tomography (QOCT). BS stands for beam splitter and $\tau$ is a
temporal delay. D$_{1}$ and D$_{2}$ are single-photon-counting
detectors that feed a coincidence circuit.}
 \end{figure}

\begin{figure}[h]
 \centering
 \epsfxsize=6.8in \epsfysize=5.2in
 \epsffile{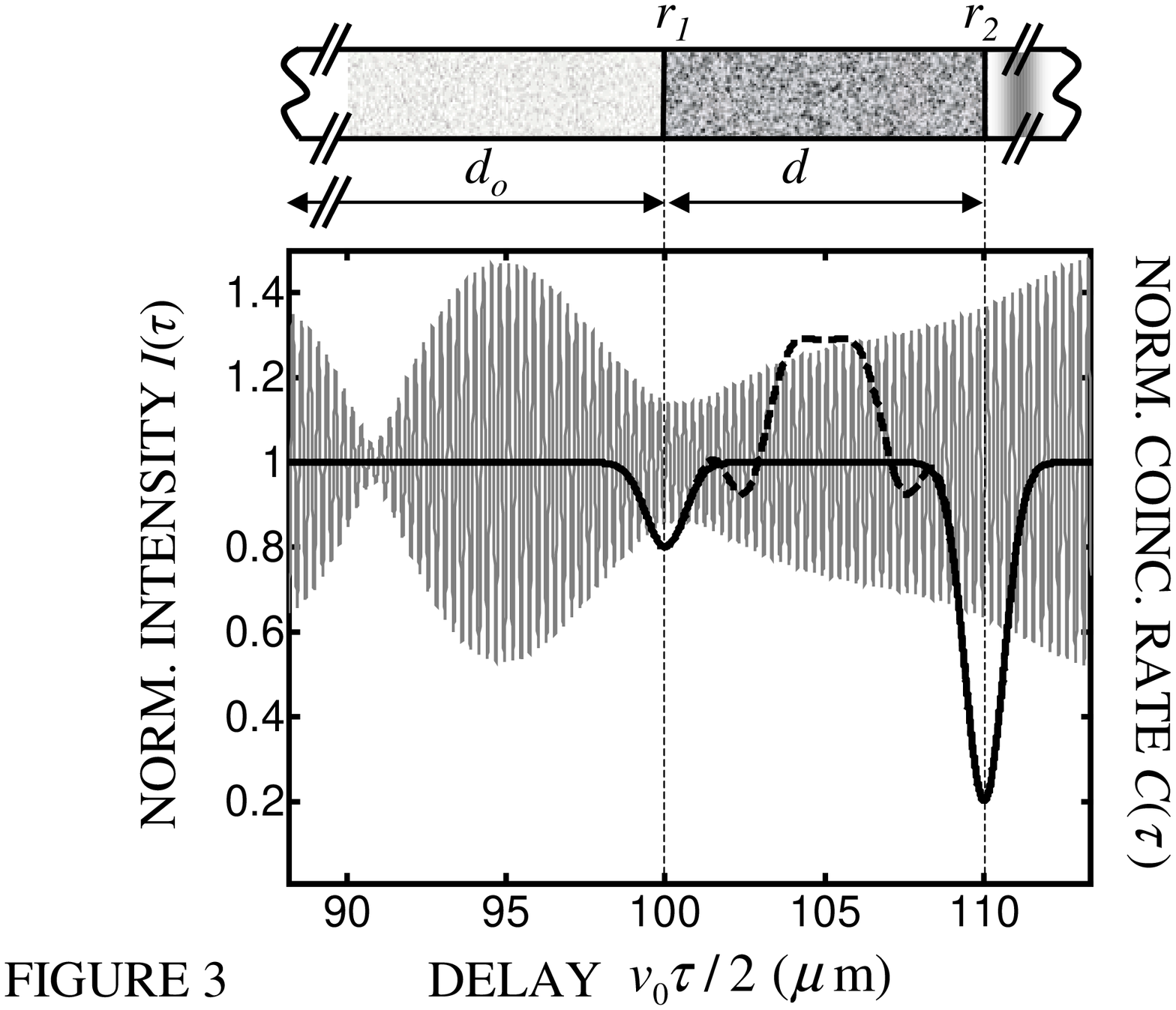}
 \vglue 0.2cm
 \label{Fig.1}\caption{Normalized intensity $I(\tau)$ (thin
rapidly varying gray curve; left ordinate) and normalized
coincidence rate $C(\tau)$ (thick black curves; right ordinate)
versus normalized delay (scaled by half the group velocity
$v_{0}/2$) for a two-layer sample buried under a dispersive
medium. The black broken curve represents the full QOCT signal
[Eq. (\ref{QOCTDiscreteSample})] whereas the black solid curve
represents the QOCT signal after averaging over the pump frequency
[Eq. (\ref{QOCTDiscreteSample}), first contribution]. The black
broken curve coincides with the black solid curve everywhere
except where the black broken curve is visible. The structure of
the sample is shown at the top of the figure. The OCT signal
yields no useful information, whereas the QOCT signal, by virtue
of the dispersion-cancellation properties of this technique,
clearly reveals the presence of the surfaces in the sample.}
\end{figure}

\begin{figure}[h]
 \centering
 \epsfxsize=6.8in \epsfysize=5.2in
 \epsffile{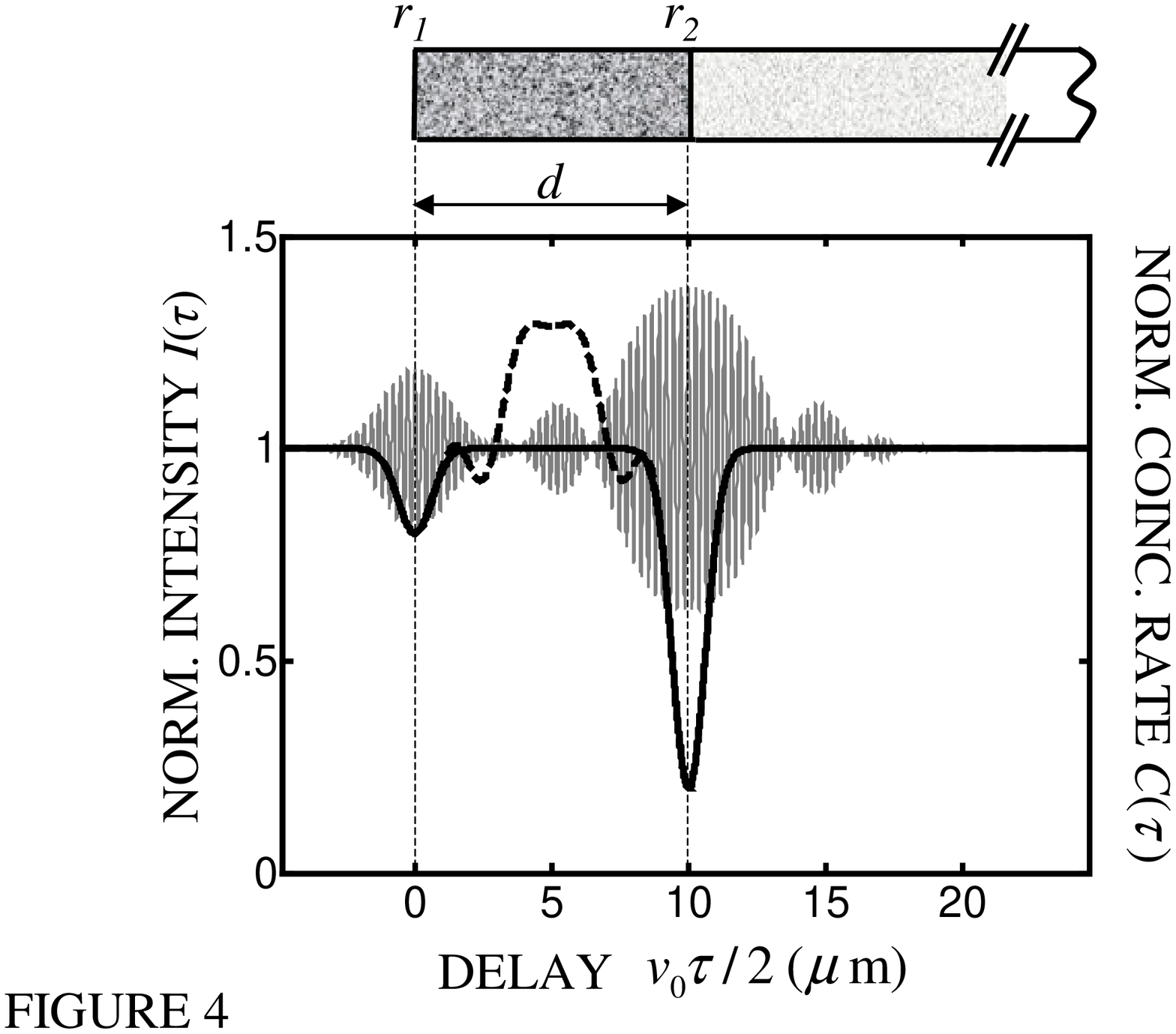}
 \vglue 0.2cm
 \label{Fig.1}\caption{Normalized intensity $I(\tau)$ (left
ordinate) and normalized coincidence rate $C(\tau)$ (right
ordinate) versus normalized delay for a two layer sample at the
surface of a medium. Curves have the same significance as in Fig.
3.}
 \end{figure}

\end{document}